\documentclass[prd,showpacs,nofootinbib,twocolumn,amsmath,amsfonts,amssymb]{revtex4} 

\usepackage{amsmath}
\usepackage{amssymb}
\usepackage{epsfig}

\usepackage{graphicx}
\usepackage{stmaryrd}
\usepackage{bm}
\usepackage{color}

\newcommand{\eps}  {\epsilon}

\newcommand{\AddrAHEP}{%
 AHEP Group, Institut de F\'{\i}sica Corpuscular --
 C.S.I.C./Universitat de Val{\`e}ncia, \\
 Edificio Institutos de Paterna, P.O. Box 22085, E--46071 Valencia, Spain}

\begin{document}
\title{Confusing nonzero $\theta_{13}$ with nonstandard interactions in the solar neutrino sector }
\author{A. Palazzo$^a$, J. W. F. Valle$^a$}
\affiliation{\AddrAHEP}

\date{\today}

\begin{abstract}
 
  Solar and KamLAND data are in slight tension when interpreted in the
  standard two-flavor oscillations framework and this may be
  alleviated allowing for a nonzero value of the mixing angle
  $\theta_{13}$.  Here we show that, likewise, nonstandard
  flavor-changing interactions (FCI), possibly intervening in the
  propagation of solar neutrinos, are equally able to alleviate this
  tension and therefore constitute a potential source of confusion in
  the determination of $\theta_{13}$.  By performing a full
  three-flavor analysis of solar and KamLAND data in presence of FCI
  we provide a quantitative description of the degeneracy existing
  between $\theta_{13}$ and the vectorial coupling
  $\eps_{e\tau}^{{\mathrm d}V}$ characterizing the nonstandard
  transitions between $\nu_e$ and $\nu_\tau$ in the forward scattering
  process with d-type quarks. We find that couplings with magnitude
  $\eps_{e\tau}^{{\mathrm d}V} \sim$ 10\%, compatible with the
  existing bounds, can mimic the nonzero values of $\theta_{13}$
  indicated by the latest analyses.

\end{abstract}


\maketitle

\section {Introduction} 

After many decades of efforts neutrino oscillations have been
definitively identified as the leading mechanism governing the flavor
transitions observed in a variety of experimental setups exploiting
both natural and artificial sources of neutrinos. In analogy with the
quark sector, where the mixing is described by the unitary Cabibbo-Kobayashi-Maskawa matrix,
the leptonic mixing is also described by a nontrivial matrix
connecting the flavor ($\nu_\alpha$, $\alpha = e, \mu, \tau$) and the
mass eigenstate neutrinos ($\nu_i, i =1,2,3$). However, it has long
been noted~\cite{Schechter:1980gr} that the matrix describing the
propagation of light neutrinos is substantially more complex than the
quark mixing matrix in that (i) deviations from unitarity may appear
and (ii) there are additional CP phases, which have no quark
analogue. Here we neglect both. Regarding the first we simply assume
that their magnitude is small, as expected in high-scale seesaw
schemes\footnote{Low-scale seesaw
  models~\cite{mohapatra:1986bd,Bazzocchi:2009kc,He:2009xd,Malinsky:2005bi,Ibanez:2009du}
  would provide an exception to this, leading to large flavor and CP
  violating effects in the charged lepton
  sector~\cite{Bernabeu:1987gr,Branco:1989bn,Deppisch:2004fa,Hirsch:2009mx,He:2009xd},
  as well as effects in neutrino
  propagation~\cite{Valle:1987gv,nunokawa:1996tg,EstebanPretel:2007yu}.
}.  It is also well-known that Majorana CP phases affect only
lepton-number violating
processes~\cite{Doi:1980yb,bilenky:1980cx,Schechter:1981gk}, and therefore can
be neglected when discussing conventional neutrino oscillations.

Hence, for simplicity, here we adopt the unitary approximation for the
lepton mixing matrix {\it U} describing neutrino oscillations, for which we
take the standard factorized parametrization given
in~\cite{Schechter:1980gr}
\begin{equation}
\label{eq:U3nu}
U = 
U(\theta_{23}) U(\theta_{13}) U(\theta_{12})
\,.
\end{equation} 
as a product of three complex rotations characterized by three mixing
angles ($\theta_{12 }$, $\theta_{13}$, $\theta_{23}$) and three
corresponding CP violating phases, and adopt the ordering prescription
of the PDG~\cite{Amsler:2008zzb}. There are two Majorana
phases which do not contribute to neutrino oscillations; the Dirac phase
is set to zero since current experiments show no sensitivity.

The small mixing angle $\theta_{13}$ is still unknown and its
measurement constitutes one of the major goals in particle physics, as
it will open the door to possible measurements of CP violation in the
leptonic sector~\cite{Bandyopadhyay:2007kx,Nunokawa:2007qh}.

Sensitivity studies of {\em future} experimental
setups~\cite{Huber:2001de,Huber:2002bi} have evidenced how the
identification of $\theta_{13}$ can be problematic due to a potential
confusion problem which may arise if nonstandard interactions (NSI)
are present. These new interactions typically arise in low-scale
models of neutrino mass, such as radiative
ones~\cite{zee:1980ai,babu:1988ki,AristizabalSierra:2006gb}, in the
form of low-energy four-fermion operators
$\mathcal{O}_{\alpha\beta}\sim \overline\nu_\alpha\nu_\beta \overline
f f$, inducing either flavor-diagonal ($\alpha = \beta$) or
flavor-changing ($\alpha \neq \beta$) neutrino transitions in the
forward scattering with the background $f$ fermion
~\cite{Wolfenstein:1977ue,Valle:1987gv,guzzo:1991hi,roulet:1991sm}. More
specifically, flavor-changing interactions (FCI) inducing transitions
among $\nu_e$ and $\nu_\tau$ ($\alpha = e$, $ \beta = \tau$) have been
recognized as an important source of confusion,
as they can mimic the effect of nonzero $\theta_{13}$ at neutrino
factories~\cite{Huber:2001de,Huber:2002bi}.

The possibility that an analogous difficulty may be {\em already} present in
the interpretation of the {\em available} neutrino data has not been
considered so far.  Indeed, in the past years the null result reported
by the short-baseline CHOOZ reactor experiment~\cite{Apollonio:2002gd}
($\sin^2\theta_{13}\lesssim $~few\%) was corroborated by the
independent findings of the global neutrino data analyses.  As a
consequence, various works aimed at establishing the (subleading) role
of NSI in the neutrino oscillation phenomenology have focused on the
well-motivated two-flavor limit, in which $\theta_{13} =0$ is
assumed. Furthermore, none of such analyses have evidenced any
significant preference for NSI.

Recently, however,  a nonzero value of $\theta_{13}$ has been hinted in various
analyses~\cite{Fogli:2008cx, Balantekin:2008zm, Fogli:2008jx,
  Fogli:2008ig, Schwetz:2008er, Escamilla:2008vq, Ge:2008sj,
  Roa:2009wp} of the latest neutrino oscillation data.  In particular,
all the existing analyses~\cite{Fogli:2008jx,Schwetz:2008er,Ge:2008sj}
find such a feature in the ``solar sector'' (solar and KamLAND data),
while its presence in the ``atmospheric sector'' (atmospheric and
$\nu_\mu \to \nu_\mu$ disappearance long-baseline data) is more
uncertain, being found in some analyses~\cite{Fogli:2005cq,
  Fogli:2008jx, Escamilla:2008vq,Roa:2009wp} but not in
others~\cite{Schwetz:2008er, Maltoni:2008ka}. Interestingly, the
preliminary searches performed by the Main Injector Neutrino
Oscillation Search (MINOS) in the $\nu_{\mu} \to \nu_e$ appearance
channel~\cite{sanchez_09} seem to support such hints, showing a weak
preference for a nonzero value of $\theta_{13}$, just below the upper
limit established by CHOOZ.

In view of these hints we deem timely to investigate whether a
confusion problem may already exist in the interpretation of the
present data. Such an issue seems even more pressing, considering that
the clearest hint of nonzero $\theta_{13}$ comes from the solar
sector, which naturally offers a sensitive setting, where the first
signs of NSI may possibly emerge.  Indeed, one should note that the
``solar'' hint of nonzero $\theta_{13}$ arises from a tension among
the standard interpretation of two-flavor transitions in matter (solar
$\nu$'s) --- where NSI may intervene --- and in vacuum (KamLAND) ---
where NSI are unimportant.  Although the standard $3\nu$
interpretation ($\theta_{13}>0$) seems the most natural one, the
possibility that this tension can be the result of some unknown effect
intervening in solar flavor transitions cannot be discarded {\em a
  priori}. Here we investigate the possibility that such an effect may
result from the theoretically well-motivated FCI, analyzing in detail
their impact on the extraction of the estimates of $\theta_{13}$ from
the presently available data.

The paper is organized as follows.  In Sec.~II, we review the notation
necessary to describe three-flavor transitions in matter in the presence
of NSI. In Sec.~III, we present the results of our numerical analysis
in the framework of $2\nu$ ($\theta_{13} = 0$) and $3\nu$
($\theta_{13} > 0$) matter transitions in the presence of FCI. We draw
our conclusions and discuss future perspectives in Sec. IV.
    
\section{Notation}

The $3\nu$ evolution in the flavor basis $(\nu_e, \nu_\mu, \nu_\tau)$ 
is described by the equation 
\begin{equation}
\label{eq:2nuevol}
 i\, \frac{d}{dx}\left(\begin{array}{c}\nu_e\\ \nu_\mu \\ \nu_\tau \end{array}\right) = H
 \left(\begin{array}{c}\nu_e\\ \nu_\mu \\ \nu_\tau \end{array}\right)\ ,
\end{equation}
 where $H$ is the total Hamiltonian,
\begin{equation}
H = H_\mathrm{kin} + H_\mathrm{dyn}^\mathrm{std} + H_\mathrm{dyn}^\mathrm{NSI}  
\label{eq:H_tot}\,,
\end{equation}
split as the sum of the kinetic term, the standard MSW (Mikheev-Smirnov-Wolfenstein)
matter term~\cite{Wolfenstein:1977ue, Mikheev:1986gs}, and of a new,
NSI-induced, matter term~\cite{Valle:1987gv}. Indicating with U the $3\times3$
mixing matrix, the kinetic term reads
\begin{equation}
\label{eq:3Hkin}
  H_\mathrm{kin} = U
  \left(
  \begin{array}{ccc}
  -\delta k/2  & 0 & 0 \\
   0 & +\delta k/2 & 0 \\
   0 &  0 &    k/2
  \end{array}
  \right)
  U^\dag\ ,
\end{equation}
where $E$ is the neutrino energy and $\delta k =\delta m^2/2E$, $k=m^2/2E$ 
($\delta m^2$ and $m^2$ being the ``solar'' and ``atmospheric'' neutrino squared mass
differences, respectively).  
The second term $H_\mathrm{dyn}^{\mathrm {std}}$ 
describes the standard (MSW) dynamics in matter~\cite{Wolfenstein:1977ue, Mikheev:1986gs},
and is given by
\begin{equation}
\label{eq:Vabstd}
  H^\mathrm{std}_\mathrm{int}=\mathrm{diag}(V,\,0,\,0)\ ,
 \end{equation}
 where $V(x) =\sqrt 2 G_F N_e(x)$ is the effective potential induced
 by the interaction with the electrons with number density $N_e(x)$.
 The term characterizing the nonstandard dynamics, assuming for definiteness 
 interactions only with d-type quarks, can be cast in the
 form
\begin{equation}
  (H_\mathrm{dyn}^{\mathrm {NSI}})_{\alpha\beta} = \sqrt{2}\,G_F\,N_d(x)\epsilon_{\alpha\beta}
\label{eq:H_NSI}\,,
\end{equation}
where $\epsilon_{\alpha\beta} \equiv  \epsilon_{\alpha\beta}^{dV}$ 
are the dimensionless vectorial couplings between neutrinos
with flavors ($\alpha, \beta$) with d-type quarks having
number density $N_d(x)$. In the phenomenological approximation 
of one-mass-scale dominance,
\begin{equation}
\label{eq:hier}
  \delta m^2 \ll m^2,
\end{equation}
we can take the limit $m^2 \to \infty $, and, similarly to the
standard MSW case~\cite{Kuo:1989qe}, reduce the $3\nu$
dynamics to an effective $2\nu$ one~\cite{Guzzo:2001mi}.  In fact, the
$3 \times 3$ mixing matrix $U(\theta_{12},\,\theta_{13},\theta_{23})$
can be factorized (assuming no CP violating phase) into three real
rotations
\begin{equation}
\label{eq:R3nu}
U \equiv R = 
R(\theta_{23}) R(\theta_{13}) R(\theta_{12})
\,.
\end{equation}
Performing a rotation of the initial neutrino (flavor) basis 
by $R^T(\theta_{13})R^T(\theta_{23})$,  and extracting the
submatrix with indices $(1,\,2)$ one finds that the survival probability 
of solar electron neutrinos is given by
\begin{equation}
\label{eq:pee3nu}
P_{ee} = c_{13}^4 P_{ee}^{\mathrm{eff}} + s_{13}^4\,,
\end{equation}
where $s_{ij} \equiv \sin \theta_{ij}$, $c_{ij} \equiv \cos \theta_{ij}$, 
and $P_{ee}^{\mathrm{eff}}$ is the $\nu_e$ survival probability in an
effective $2 \times 2$ model described by the Hamiltonian
\begin{equation}
    H^{\mathrm{eff}} = V(x)
    \begin{pmatrix}
        c^2_{13} & 0 \\
        0        & 0
    \end{pmatrix}
    + \sqrt 2 G_f N_d(x)
    \begin{pmatrix}
        0         & \eps \\
        \eps  & \eps'
    \end{pmatrix}
\,,    
\end{equation}
where $\eps$ and $\eps'$  are two effective parameters which,
considering only FCI [$\alpha \ne \beta$ in Eq.~(\ref{eq:H_NSI})], are 
related to the original $\eps_{\alpha\beta}$ couplings,  as~\cite{Guzzo:2001mi}
\begin{align}
    \label{eq:4} \eps
    &= c_{13} ( \eps_{e\mu} c_{23} - \eps_{e\tau} s_{23} )
    - s_{13} [ \eps_{\mu\tau} (c_{23}^2 - s_{23}^2)] \,,
    \\
    \begin{split}
        \label{eq:5} \eps'
        &= - 2 \eps_{\mu\tau} c_{23} s_{23}
        + 2 s_{13} c_{13} (\eps_{e\tau} c_{23} + \eps_{e\mu} s_{23}) \\
        &\quad
        - 2 s_{13}^2 \eps_{\mu\tau} s_{23} c_{23} \,.
    \end{split}
\end{align}
Considering we are interested in only the FCI  between
$\nu_e$ and $\nu_\tau$,
we then remain with the expressions
\begin{align}
    \label{eq:4a} \eps
    &= - \eps_{e\tau} c_{13}  s_{23}\,,
    \\
 \begin{split}
        \label{eq:5a} \eps'
        &= 
        + 2 \eps_{e\tau} s_{13} c_{13} c_{23}\, 
       \end{split} \,.
\end{align}
Therefore the propagation of solar neutrinos can be described
effectively by a two-dimensional evolution Hamiltonian, which depends
on the five parameters ($\delta m^2, \theta_{12}, \theta_{13},
\theta_{23}, \eps_{e\tau}$).

\section{Numerical results}

In our analysis we have included the data from the radiochemical
experiments Homestake~\cite{cleveland:1998nv},
SAGE~\cite{abdurashitov:2002nt} and
GALLEX/GNO~\cite{Hampel:1998xg,Altmann:2005ix,kirsten2008retrospect},
Super-KamioKande~\cite{fukuda},
from all the three phases of the Sudbury Neutrino Observatory
(SNO)~\cite{ahmad,Ahmed:2003kj,Aharmim:2005gt,Aharmim:2008kc},
and Borexino~\cite{Arpesella:2008mt}. We have also included the latest
KamLAND data~\cite{:2008ee}. For the sake of precision, we have
incorporated both standard and nonstandard matter effects in
KamLAND. However, due to the low density of the Earth's crust,
both have only a negligible effect for the range of parameters we are
considering.  Therefore, the constraints obtained on KamLAND do not
depend on NSI. We also included NSI effects in the
propagation of solar neutrinos in the Earth which, as noted
in~\cite{Friedland:2004pp}, can modify the regeneration effect.

We begin our study considering the more familiar two-flavor case
($\theta_{13} = 0$) in which the results of our analysis depend on the
three parameters: ($\delta m^2, \theta_{12}, \eps_{e\tau}$).  Indeed,
we can safely assume $\sin^2\theta_{23} = 1/2 $, motivated by the
results of the latest atmospheric neutrino data analyses, which
indicate maximal~\cite{Schwetz:2008er} or nearly
maximal~\cite{Fogli:2008ig, GonzalezGarcia:2007ib} mixing.  In Fig.~1
we show the region allowed by KamLAND (horizontally elongated region)
in the plane spanned by the standard oscillation parameters [$\delta
m^2, \sin^2\theta_{12}$], superimposed to the solar large mixing angle (LMA)
region of oscillation parameters obtained in the absence of FCI
($\eps_{e\tau} = 0$), and for two representative cases in which FCI
are ``switched on,'' with couplings having the same amplitude but
opposite sign ($\eps_{e\tau} = \pm 0.2$).
\begin{figure}[t!]
\vspace*{-2.5cm}
\hspace*{0.5cm}
\includegraphics[width=10 cm]{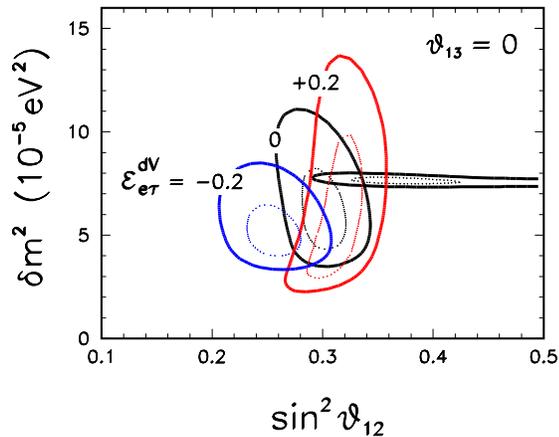}
\vspace*{-1.7cm}
\caption{The solar LMA region is represented  at two C.L.'s [$\Delta \chi^2 = 1$ (thin dashed line)
and $\Delta \chi^2 = 4$ (thick solid line)] for the standard case 
($\eps_{e\tau}^{dV} = 0$) and for two representative nonstandard cases
with FCI having equal amplitude and opposite sign ($\eps_{e\tau}^{dV} = \pm 0.2$).
The horizontally elongated regions are those allowed by KamLAND
(at the same C.L's.) which do not depend on FCI.}
\end{figure}  
The most important clue we have from this plot is the shift of the
solar LMA region in the horizontal direction in correspondence of the
(phenomenologically relevant) values of $\delta m^2$ determined by
KamLAND.  Indeed, the solar LMA region ``moves'' towards higher
(lower) values of $\theta_{12}$, for positive (negative) values of
$\eps_{e\tau}$. From Fig.~1, it is clear that positive values of
$\eps_{e\tau}$ tend to reduce the tension between solar and KamLAND
and are preferred in the solar+KamLAND combination, which gives
$\eps_{e\tau} \simeq 0.15$ as best fit and disfavors the standard case
($\eps_{e\tau} =0$) at $\sim 1.3 \sigma$ level ($\Delta \chi^2 \sim
1.7$).

These results strongly suggest that positive values of $\eps_{e\tau}$, 
{with size compatible with current limits~\cite{limits}, can mimic the effect of
nonzero $\theta_{13}$. Notice, however, that in the standard $3\nu$
analysis {\em both} the solar LMA region and the region allowed by 
KamLAND in the plane [$\delta m^2, \sin^2\theta_{12}$] get
modified by values of $\theta_{13} >0$, for which they
tend to merge~\cite{Fogli:2008cx,Balantekin:2008zm,Fogli:2008jx,Schwetz:2008er,Ge:2008sj}.
In contrast, in the case under consideration, only the solar LMA
region is affected.

In order to trace more quantitative conclusions it is necessary to perform a
full $3\nu$ + FCI analysis, where both $\theta_{13}$ and
$\eps_{e\tau}$ are allowed to assume nonzero values.
\begin{figure}[t!]
\vspace*{-2.2cm}
\hspace*{-0.5cm}
\includegraphics[width=12cm]{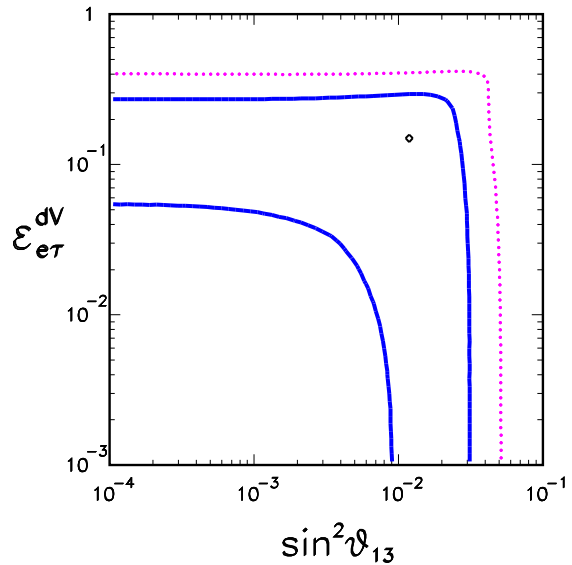}
\vspace*{-3.0cm}
\caption{Region allowed by the combination of solar and KamLAND data
at two C.L`s  [$\Delta \chi^2 =1$ (solid line) and  $\Delta \chi^2 =4$ (dashed line)] 
after marginalization of $\delta m^2$ and $\theta_{12}$.
\label{fig2}}
\end{figure}  
In Fig.~2 we display the main result of such an analysis, by showing
the constraints (at $\Delta \chi^2 =1, 4$) obtained from the
combination of solar and KamLAND, in the plane of the two relevant
parameters [$\sin^2\theta_{13}, \eps_{e\tau}$], after marginalization
over the remaining parameters $\delta m^2$ and $\sin^2\theta_{12}$.  We display only the
region corresponding to positive values of
$\eps_{e\tau}$, which are relevant for the degeneracy problem under
study.  One sees that, at low confidence levels the
preferred region is a band (delimited by the solid curves) which does not contain
the origin (``disfavored'' at $\Delta \chi^2 \simeq 2.0$). The
slight tension among solar and KamLAND gets effectively diluted among
the two parameters $\theta_{13}$ and $\eps_{e\tau}$. For higher
confidence levels (dotted curve), the allowed region does contain the origin and only
upper limits can be put on both parameters.  These results allow us
to conclude that a complete degeneracy between the two parameters is
present in the current neutrino data.

The question arises as to whether and how future data may remove such
a degeneracy.  With this purpose, in Fig.~3 we show the behavior of
the solar $\nu_e$ survival probability (averaged over the $^8${\it B}
$\nu$ production region) profile $P_{ee} (E)$, for three
representative cases.  In all of the three cases presented the
probability is calculated for the fixed values of the leading
parameters ($\delta m^2 = 7.67\times10^{-5} \mathrm{eV}^2$,
$\sin^2\theta_{12} = 0.3$).  The three curves correspond to the
following cases: (I) The solid line represents the case of pure 2$\nu$
standard transitions ($\theta_{13} = 0$, $\eps_{e\tau} = 0$),
corresponding to the origin in Fig.~2; (II) The dashed line indicates a
representative case of standard 3$\nu$ transitions ($\sin^2\theta_{13} =
0.02$, $\eps_{e\tau} = 0$); (III) The dotted line shows a
representative case of $2\nu$ + FCI transitions ($\theta_{13} = 0$,
$\eps_{e\tau} = 0.1$).

As one can infer from Fig.~2, cases (II) and (III) have been chosen
so as to correspond to (currently) indistinguishable points in
parameter space.  With respect to case (I), regarded as a benchmark,
we note the following differences between the two degenerate cases, (II)
and (III).  In the standard $3\nu$ case (dashed line) $P_{ee}$ is
suppressed with respect to the standard $2\nu$ case (solid line) by the
energy independent factor $\sim 1-2\sin^2\theta_{13}$ [see
Eq.~(\ref{eq:pee3nu})].  In contrast, the $2\nu$ + FCI case (dotted
line), is characterized by an energy-dependent suppression%
\footnote{A similar behavior has been noticed in~\cite{Friedland:2004pp}.} 
respect to the standard $2\nu$ case. In
particular, the suppression is completely negligible at low energies
($E<3~$MeV), and it is more pronounced at intermediate energies.  The
net effect is a flattening of $P_{ee}(E)$ with an enhancement of the
up-turn typical of the adiabatic MSW transitions.  

The current data
are unable to distinguish between case (II) and (III) since: (i) the
differences at low energies are too small to be detected by the
gallium experiments or Borexino; (ii) the current high energy
experiments are not sensitive enough to probe the up-turn region,
which still remains practically ``invisible.''  The differences at low
energy between the two degenerate cases, (II) and (III), are very tiny and
may prove very hard to detect even at future low-energy experiments.
Instead, the possibility to disentangle the different behavior at
intermediate energies could perhaps become realistic in high-energy
experiments with a lowered threshold. In this respect, the new data
expected from Borexino, Super-K-III~\cite{Takeuchi:2008zz}, and from
the low energy threshold analysis underway in the SNO
collaboration~\cite{SNO-LETA}, may play an important role.

\begin{figure}[t!]
\vspace*{-1.0cm}
\hspace*{0.2cm}
\includegraphics[width=6.7 cm]{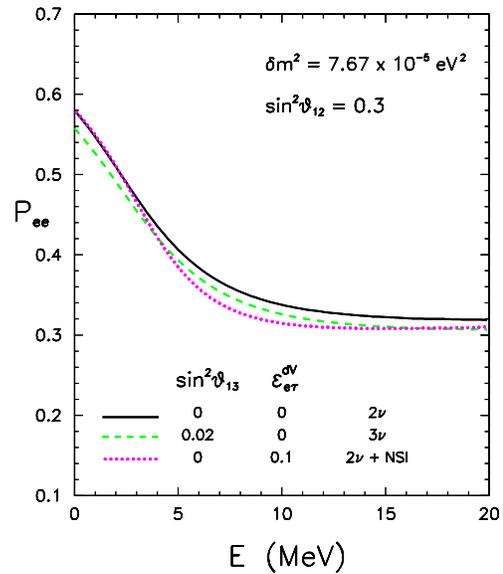}
\vspace*{-0.0cm}
\caption{Solar $\nu_e$ survival probability (averaged over the $^8B$ $\nu$
production region) for three representative cases.}
\end{figure}  
We close this section with a final remark.  For definiteness, we have
focused on the case of interactions with d-type quarks and
transitions among $\nu_e$ and $\nu_\tau$.  However, the essence of
our conclusions is unaltered if interactions with u-type quarks or
electrons and/or transitions among $\nu_e$ and $\nu_\mu$ ($\alpha = e,
\beta = \mu $) are considered.  Moreover, the simultaneous 
inclusion of more than one type of NSI  can only exacerbate the
confusion problem we have posed, leading to further difficulties in data
interpretation.

\section{Conclusions}

We have stressed that nonstandard flavor-changing interactions may
constitute a source of confusion in the interpretation of {\em
  present} solar and KamLAND neutrino data, hindering the correct
determination of the mixing angle $\theta_{13}$. In the near future,
various solar experiments may help to reduce such a difficulty by
providing a more precise determination of the energy profile of the
solar $\nu_e$ survival probability.  In the mean time, it would be
very important to complement our study --- focused on the solar sector
--- with a similar investigation on the atmospheric sector, which
could hopefully be of aid in breaking the degeneracy among
$\theta_{13}$ and $\eps_{e\tau}$. Also, a quantitative assessment of
the impact of nonstandard neutrino interactions in the interpretation
of the preliminary MINOS data in the $\nu_\mu \to \nu_e$ appearance
channel~\cite{sanchez_09} would be highly desirable.  Our results
underline the importance of a ``clean'' measurement of $\theta_{13}$
expected from the new generation reactor
experiments~\cite{reactor}, whose inferences are
free from NSI effects.  We also stress how, in the event of a null
result by these experiments (i.e., in the case of nonconfirmation of
the present hints of nonzero $\theta_{13}$), a persisting tension
among solar and KamLAND would pose a novel problem, whose resolution
may involve the nonstandard interactions discussed here or other
possible unaccounted effects.


\section*{Acknowledgments}
We would like to thank
M.A. T\'ortola for precious discussions. 
This work was supported by the Spanish Grants No. FPA2008-00319/FPA,
No. FPA2008-01935-E/FPA and No. PROMETEO/2009/091.  A.P.\ is supported by MEC
under the I3P program.

\bibliographystyle{h-physrev4}

\end{document}